\documentclass[12pt]{article}

\usepackage{scicite}
\usepackage{graphicx}
\usepackage{times}

\newenvironment{sciabstract}{%
\begin{quote} \bf}
{\end{quote}}

\newcounter{lastnote}

\title{Spin-Triplet Superconductivity in K$_{2}$Cr$_{3}$As$_{3}$} 

\author
{Jie Yang,$^{1}$ Jun Luo,$^{1}$ Changjiang Yi,$^{1}$ Youguo Shi,$^{1}$ Yi Zhou,$^{1,2}$ Guo-qing Zheng$^{3\ast}$ \\
\\
\normalsize{$^{1}$Institute of Physics, Chinese Academy of Sciences and}\\
\normalsize{Beijing National Laboratory for Condensed Matter Physics, Beijing 100190, China}\\
\normalsize{$^{2}$Kavli Institute for Theoretical Sciences, CAS Center for Excellence in Topological Quantum Computation,}\\ \normalsize{University of Chinese Academy of Sciences, Beijing 100190, China}\\
\normalsize{$^{3}$Department of Physics, Okayama University, Okayama 700-8530, Japan}\\
\\
\normalsize{$^\ast$To whom correspondence should be addressed; E-mail:  zheng@psun.phys.okayama-u.ac.jp}
}

\date{}

\begin{document}


\baselineskip24pt


\maketitle

\clearpage
\begin{sciabstract}

A spin-triplet superconductor can harbor Majorana bound states that can be used in topological quantum computing.
Recently,  K$_{2}$Cr$_{3}$As$_{3}$ and its variants with critical temperature $T_{\rm c}$ as high as 8 K have emerged as a new class of superconductors with ferromagnetic spin fluctuations. Here we report a discovery in K$_{2}$Cr$_{3}$As$_{3}$ single crystal that, the spin susceptibility measured by $^{75}$As Knight shift below $T_{\rm c}$ is unchanged with the magnetic field $H_{\rm 0}$ applied in the  $ab$ plane, but vanishes toward zero temperature when $H_{\rm 0}$ is along the $c$ axis, which unambiguously establishes this compound as a spin-triplet superconductor described by a vector order-parameter   $\vec{d}$
 parallel to the $c$ axis.
 Combining with points-nodal gap we show that  K$_{2}$Cr$_{3}$As$_{3}$ is a new platform for the study of topological superconductivity and its possible technical application.

\end{sciabstract}

\textbf{Teaser:}
A spin-triplet superconductor with ferromagnetic spin fluctuation superconducts at a critical temperature of 6.5 K


\section*{Introduction}
In conventional superconductors, the electron pairs (Cooper pairs) are bound by the electron-phonon interaction,
which results in a superconducting state with symmetric  ($s$-wave) orbital wave function and antisymmetric spin orientation (spin singlet, $S$=0) \cite{BCS}.
In the cuprate high temperature superconductors, Cooper pairs are also in a spin-singlet state but with $d$-wave symmetry for the orbital wave function, which are believed to be mediated by  antiferromagnetic spin fluctuations \cite{cuprates}.
In superfluid $^3$He, however, Cooper pairs are in spin triplet state ($S$=1) with $p$-wave orbital wave function that is antisymmetric  about the origin (odd parity) \cite{Leggett}. In this case, the spin triplet state is favored by  ferromagnetic (FM) spin fluctuations.
Spin-triplet $p$-wave superfluid state is also believed to be realized in neutron stars \cite{neutronstars}.
For long time, a solid-state analogue of $^3$He 
has been  sought 
in 
strongly correlated electron
systems (SCES) where the superconducting transition temperature $T_{\rm c}$ is typically around 1 K \cite{SrRuO,UPt3Review,UGe2,URhGe,UCoGe,UTe2}, 
but  unambiguous evidence is still lacking. 
The most promising SCES candidate had been Sr$_{2}$RuO$_{4}$  ~\cite{SrRuO,Rice}, but recent experimental progress ~\cite{Brown} has casted doubt on its pairing symmetry.

The  spin triplet state that requires  odd-parity for the orbital part of wave function is particularly fascinating and important from the topological point of view.
%
 This is because an odd-parity superconductor can be topological \cite{Fu_Berg_PhysRevLett.105.097001,Matano} and can host 
 Majorana bound states in their vortex cores or chiral Majorana fermions on boundary (surface or edge)  \cite{Wilczek}, which are robust against scattering. Thus, spin-triplet superconductors are  of great  interests and importance not only in fundamental physics~\cite{Qi}, but also  in applications as their edge or bound states  can be used to implement topological quantum computing based on  non-Abelian 
 statistics~\cite{Kitaev}. 
 Thus far,
efforts of questing topological superconductivity have been devoted in two directions; one is via exploring bulk materials and the other is through utilizing surface states often induced by proximity effect. 
Physical probes aiming 
to identify such novel state can also be divided into two  categories: bulk properties or edge states measurements.
 Although  some progress  has been made by surface-sensitive probes in looking for signatures of  edge  states  
  due to superconducting proximity effect
  ~\cite{Fu,Mourik,Jiajinfeng,Ding}, 
searching for intrinsic topological superconductivity in bulk materials is highly desired.
%
In addition to the case of odd parity, if a  superconducting state breaks time reversal symmetry, the superconductivity is topological \cite{Senthil}. Superconductivity in a crystal that breaks spatial inversion symmetry has also a good chance to be topological \cite{sato_PhysRevB.79.094504,Li2Pt3B}.  

Recently, a new superconducting  family containing 3d transition-metal element  Cr,  A$_{2}$Cr$_{3}$As$_{3}$ (A = Na, K, Rb, Cs) has  been reported~\cite{K, Rb, Cs, Na}, with a $T_{\rm c}$ as high as 8 K. Signatures for   unconventional superconductivity have been  found ~\cite{Imai,Canfield,Yuan_HQ,usr,Yang,LuoPRL}. 
Nuclear magnetic resonance (NMR) measurements reveal point nodes in the superconducting gap function \cite{Yang,LuoPRL} and ferromagnetic  spin fluctuation in the normal state 
which can be tuned by changing the alkali ion radius 
\cite{LuoPRL}. Thus, A$_{2}$Cr$_{3}$As$_{3}$  bear  some similarities  of superfluid $^3$He where ferromagnetic fluctuation promotes spin-triplet pairing. 


The spin susceptibility in the superconducting state is a bulk property, and that measured by the Knight shift is sensitive to  spin pairing symmetry.  
In this work, through $^{75}$As Knight shift measurements in a single crystal, 
we show that  spin nematicity (broken rotational symmetry)
spontaneously emerges    
in the superconducting state of K$_{2}$Cr$_{3}$As$_{3}$ with $T_{\rm c}$ = 6.5 K, which is a hallmark for   a spin-triplet state.
We identify the direction of the vector order parameter that describes the spin-triplet state and estimate the strength of the interaction that pinned the vector to a specific crystal axis.
We show that 
A$_{2}$Cr$_{3}$As$_{3}$
 is a new route to studying topological superconductivity and future  technical implementation using a topological spin-triplet superconductor at a high temperature.

\section*{Results}
\begin{figure}[htbp]
\includegraphics[width=14cm]{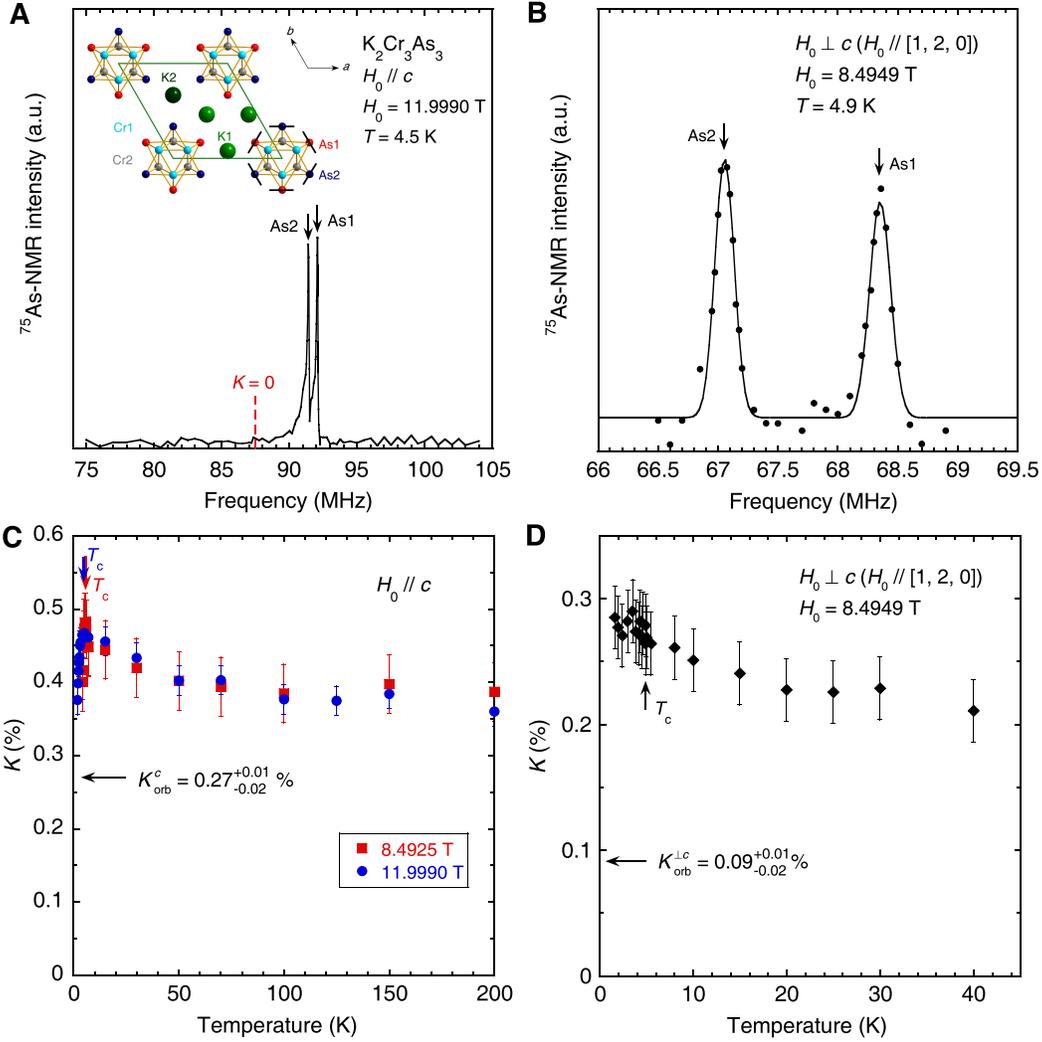}
\centering
\caption{\textbf{NMR spectra of K$_{2}$Cr$_{3}$As$_{3}$ and the obtained temperature dependence of the Knight shift for the magnetic field applied along the $c$ axis and in the $ab$ plane.} (\textbf{A, B}), The $^{75}$As-NMR spectra for $H_{0}$ $||$ $c$ axis,  and  $H_{0}$ $||$  $ab$ plane ($H_{0}$ $||$ [1,2,0]) 
	at representative temperatures. The inset to (\textbf{A}) is the top view of the crystal structure  of K$_{2}$Cr$_{3}$As$_{3}$. The green frame indicates the unit cell. There are two inequivalent As sites in the crystal lattice, i.e., As1 and As2. The principal axes of the EFG at As nuclei lie in the $ab$ plane, which are  indicated by the black bars.
(\textbf{C, D}), The temperature dependence of the Knight shift. 
The vertical arrows indicate $T_{\rm c}$ under various fields, 
and the horizontal arrow indicate the value of Knight shift due to orbital susceptibility.
The error bar for $K$
was estimated by assuming that the spectrum-peak uncertainty equals the
point (frequency) interval in measuring the NMR spectra.
\label{spectra}}
\end{figure}

{\bf NMR spectra and determination of the Knight shift.}
We performed $^{75}$As NMR measurements on a high-quality single crystal K$_{2}$Cr$_{3}$As$_{3}$ with the magnetic field $H_{\rm 0}$  applied along different  directions,  covering both the superconducting and normal states.
Figure~\ref{spectra}A and B
show representative frequency-swept $^{75}$As NMR spectra of the central transition ($I_z$ = -1/2 $\leftrightarrow$  +1/2) with   $H_{\rm 0}$ $\parallel$ $c$ axis and $H_{\rm 0}$ $\parallel$ $ab$ plane, respectively. For axially symmetric electric field gradient (EFG),  
the  transition frequency $\nu$ of the central transition 
can be written as~\cite{Abragam}
\begin{equation}
\nu = (1+K)\gamma H_{0} + \frac{{3\nu _Q^2}}{{16(1 + K){{(\gamma H_{0})}}}} {\rm sin}^2 \theta (1 - 9 {\rm cos}^2 \theta)
\label{eq1}
\end{equation}
where $K$ is the Knight shift, $\nu _Q$ is the nuclear quadrupole resonance (NQR) frequency, and  $\theta$ is the angle between $H_{\rm 0}$ and the principal  axis of the EFG at the As nucleus position. For a complete and general expression in the presence of EFG asymmetry $\eta$, see Supplementary
Materials (SM).
First-principle calculation reveals that  the principal EFG axis lies in the $ab$ plane, but is along different directions for  the six positions of As1 and As2  as shown in the inset of Fig.~\ref{spectra}A ~\cite{LuoCPB}. The obtained  $\eta$ is tiny ($\eta$=0.004) so that the correction to eq. (1) is negligible (see SM).  In our measurements,  $\theta$ = 90$^{\circ}$ when $H_{\rm 0}$ $||$  $c$.
For the measurements with $H_{\rm 0}$ $||$ $ab$,   $\theta$ is also 90$^{\circ}$  when  $H_{\rm 0}$ is parallel to the mirror planes indexed by (2,-1,0), (1,1,0) and (1,-2,0). 
Such field direction can also be described by the crystal direction indices of [1,2,0], [-1,1,0] and [2,1,0].
We obtain the the Knight shift $K$ by two different methods and the results agree well. One is based on eq. (1) with the $\nu _Q(T)$ value obtained from NQR measurements \cite{LuoPRL,Imai}, and the other is by changing the magnetic field  so that the obtained $K$ does not depend on whether $\eta$ is finite or not (see SM for details). 
The temperature variation of $K^{c}$ for $H_{\rm 0}$ $||$ $c$ axis, and $K^{\perp c}$ for $H_{\rm 0}$  $||$ $ab$  ($H_0\parallel$ [1,2,0] and equivalents) 
is shown in Fig.~\ref{spectra}C and D, respectively. 

\begin{figure}[htbp]
\includegraphics[width=16cm]{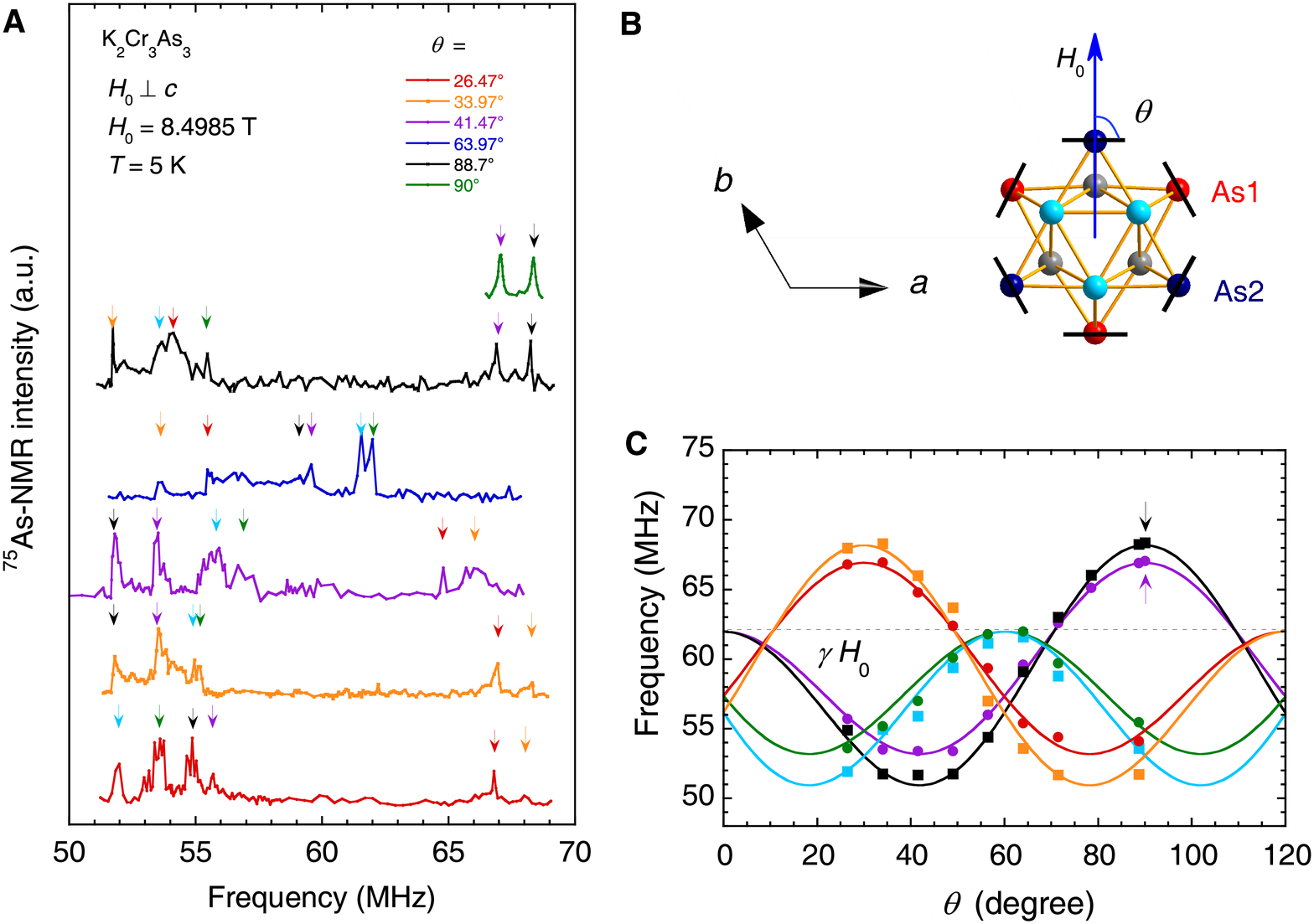}
\centering
\caption{\textbf{Angle dependent $^{75}$As NMR spectra for $H_{\rm 0}$ $||$  $ab$.} (\textbf{A}) Angle dependence of the complete $^{75}$As NMR spectra for $H_{\rm 0}$ $||$  $ab$. The peaks are marked by the arrows with six different colors, with the same color meaning that they come from the same As position. 
(\textbf{B}) Top view of the Cr-As chains of  K$_{2}$Cr$_{3}$As$_{3}$. The black bars indicate the directions of the EFG principal axes of the As sites. (\textbf{C}) Angle dependence of the  $^{75}$As central transition peak frequency. The curves are the theoretical calculation for the six As positions. The arrows indicate the sites and field directions at which the temperature dependence of the Knight shift was measured. 
\label{fullspectra}}
\end{figure}

Before proceeding further, we comment on some aspects of the spectra. Due to the crystal symmetry, 
 the EFG principal axis differs by 60$^\circ$ between the six As positions in the plane, 
so the spectra with $H_{\rm 0}$ $||$ $ab$ should be 6-fold rotation symmetric when $H_{\rm 0}$ rotates in the   $ab$ plane. Indeed, we have  directly confirmed this property.  
Figure~\ref{fullspectra} shows some representative spectra for different angle between $H_{\rm 0}$ and  the $a$-axis of the crystal. 
Since As2 site has a smaller  $\nu _{Q}$  compared to As1~\cite{LuoCPB},
the central transition has two peaks and the left peak is assigned to the As2 site while the right peak to the As1 site (Fig.~\ref{spectra}A and B).
Since the As1 and As2 sites have different $\nu_Q$ values, the central transition of $^{75}$As NMR spectrum with $H_{\rm 0}$ $||$  $ab$ will split into six peaks due to different $\theta$ for each As position, which was indeed observed as shown in Fig. ~\ref{fullspectra}A. When rotating $H_{\rm 0}$ within the $ab$ plane, the angle dependence of the spectrum peak is 6-fold  symmetric, as shown in Fig. ~\ref{fullspectra}C.   
  The observed peak positions are in good  agreement with those calculated  from Eq. (1), with $\nu_Q$ value  taken from ref. \cite{Imai,LuoPRL}.
The two sites As1 and As2 show  basically the same properties as found in previous NMR measurements~\cite{Yang,Imai} and also in the present work (see the SM for details). We therefore   focus on the As2 site hereafter.

\begin{figure}[htbp]
\includegraphics[width=12cm]{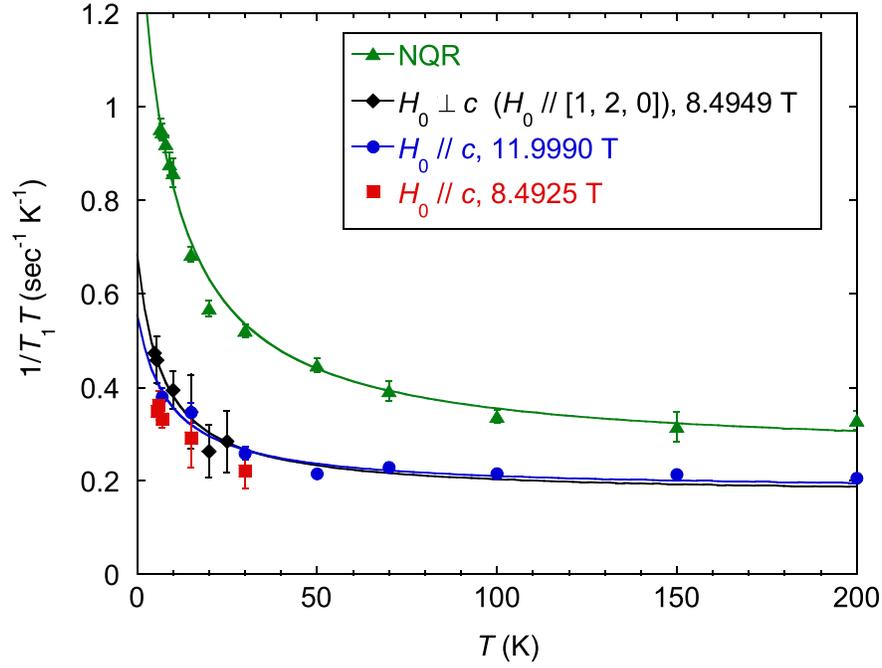}
\centering
\caption{\textbf{Temperature dependence of    1/$T_{1}T$  for As2 site in the normal state.} The 1/$T_{1}T$ increases with decreasing temperature due to the development of FM spin fluctuations.
The NQR data are taken from ref. \cite{LuoPRL}. The
	solid curves are fittings to $1/T_{1}T$ = $(1/T_{1}T)_{\rm DOS}$ + $b/(T+\theta)$. The error bar for 1/$T_{1}T$ is the standard
deviation in fitting the nuclear magnetization recovery curve.
\label{T1T}}
\end{figure}

{\bf Electron correlations.}
We then discuss the properties of the electron correlations based on the results of $^{75}$As Knight shift and  spin-lattice relaxation rate $1/T_{1}$. 
For both field orientations, the Knight shift $K$  increases with decreasing temperature below $T$ = 50 K.
We also measured 
$1/T_{1}$  
for the As2 site.
Figure~\ref{T1T}   shows  1/$T_{1}T$ above $T_{\rm c}$ as a function of temperature.
For each field orientation, $1/T_{1}T$ increases upon cooling also below $T$ = 50 K.
These results demonstrate that FM spin  fluctuations develop at low temperatures, being consistent with the previous results obtained in polycrystalline samples~\cite{Yang,LuoPRL}.

 The absolute  value of 1/$T_{1}T$ obtained by NQR is larger than that obtained in a single crystal  with $H_0\perp c$ ($H_0\parallel$ [1,2,0]) 
 or $H_0\parallel c$. This is because in the NQR measurements, the  effective $H_0$ direction is along the principal axis, which is perpendicular to both the $c$-axis and the [1,2,0] direction. 
As the hyperfine coupling is anisotropic in general, the absolute value of $1/T_1T$ along different direction can be different. As we show below, however, the temperature dependence of $1/T_1T$ for all field directions is identical.
%
In general, 1/$T_{1}T$ is proportional to the $q$-summed imaginary part of transverse dynamical susceptibility ${\chi }^{''}_{\perp }$ divided by $\omega$,
\begin{equation}
\frac{1}{{{T_1}T}} \propto \sum_q {A(q)^2}\frac{{{\chi }^{''}_{\perp }(q,\omega )}}{\omega }
\label{T1}
\end{equation}
where  \textbf{A(q)} is the hyperfine coupling tensor and $\omega$ is Larmor frequency.
When there is a peak in a specific $q=Q$ 
due to electron correlations, $1/T_1T$ may be decomposed into two parts,
\begin{equation}
1/T_{1}T = (1/T_{1}T)_{\rm DOS} +(1/T_{1}T)_{Q}
\label{T1dos}
\end{equation}
The first term is due to non-correlated electrons, being determined by the density of states (DOS) at the Fermi level,
which is usually constant. The second term is due to the development of  ferromagnetic spin fluctuation in the present case ($Q$=0). According to Moriya's theory for a  ferromagnetically correlated 3D metal \cite{Moriya}, 1/$T_{1}T$  follows a Curie-Weiss $T$-dependence  as
\begin{equation}
(1/T_{1}T)_{Q} = b/(T+\theta)
\label{T1Q}
\end{equation}
Figure~\ref{T1T} shows the fittings of $1/T_{1}T$ to Eq. (\ref{T1dos}), which reveals that the single-crystal data can be fitted by using $(1/T_{1}T)_{\rm DOS}^c$ =  $(1/T_{1}T)_{\rm DOS}^{\perp c}$ ($H_0\parallel$ [1,2,0])  = 0.18 sec$^{-1}$K$^{-1}$, and with the same  $\theta \sim$ 10 K obtained from the NQR data \cite{LuoPRL}.

{\bf Separating various contributions to the Knight shift.}
Next we discuss the various contributions to the Knight shift in the superconducting state.
The Knight shift $K$ consists of three parts,
$K = K_{\rm s} + K_{\rm orb} + K_{\rm dia}$,
where $K_{\rm s}$ is proportional to the spin susceptibility $\chi_{\rm s}$, $K_{\rm orb}$ is the contribution from orbital susceptibility and is temperature independent, and $K_{\rm dia}$ arises from  diamagnetism 
due to vortex lattice formation in the superconducting state. 
 The $K_{\rm dia}$ is calculated to be negligible in K$_{2}$Cr$_{3}$As$_{3}$ because of a large penetration depth (see SM).
 The $K_{\rm orb}$ was determined by an analysis utilizing  the relationship between $K$  and 1/$T_1T$,   and its value is respectively indicated by the horizontal arrow in Fig.~\ref{spectra}C and D. 

\begin{figure}[htbp]
\includegraphics[width=16cm]{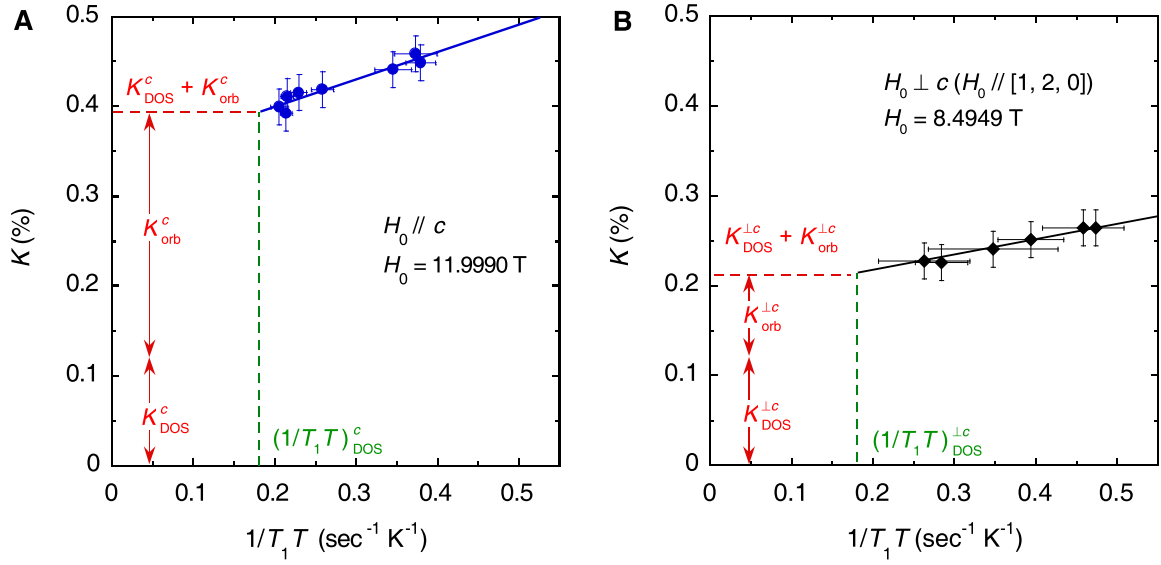}
\centering
\caption{\textbf{Determination of  the Knight shift due to orbital susceptibility ($K_{\rm orb}$).}
(\textbf{A, B}) The plot of $^{75}$As Knight shift  against $1/T_{1}T$ for $H_{0}$ $||$ $c$ axis and  $H_0 \perp c$, 
respectively. The uncertainty for  $K_{\rm orb}$ is +0.01\%/-0.02\%. 
\label{Korb}}
\end{figure}

In the following we elaborate how the $K_{\rm s}$ and $K_{\rm orb}$ are separated.
 We first note that  $K_{\rm s}$ can further be decomposed into two parts, with the first part $K_{\rm DOS}$  due to non-interacting electrons, and the second part due to $d$-electrons,  $K_{\rm s}^{\rm int}$,  which is $T$-dependent. As described above, the interacting $d$-electrons are responsible for FM fluctuation and contribute to the Curie-Weiss behaviour of $1/T_1T$ which is proportional to $\chi (\emph{\textbf{q}} = 0)$.
In such  FM spin fluctuation case, 
 $K_{s}^{\rm int}$ is also proportional to $\chi (\emph{\textbf{q}}  = 0)$~\cite{Moriya}.
Figure~\ref{Korb}  
shows the  $K$ vs. $1/T_{1}T$ plots with $H_{0}$ $||$ $c$  for  $T_c(H)$ = 5.1 K $\leq T \leq$ 200 K, and with  $H_0 \perp c$ ($H_0\parallel$ [1,2,0]) 
for $T_c(H)$ = 4.9 K $\leq T \leq$ 25 K, respectively. In both cases,
a fairly good linear relation is indeed  found, reflecting the relationship described above. The vertical dashed line indicates the position of
$(1/T_{1}T)_{\rm DOS}$ = 0.18 sec$^{-1}$K$^{-1}$ obtained from Fig. \ref{T1T} (see preceding subsection).
The corresponding $K$ indicated by the horizontal dotted line is then  $K_{\rm DOS}$+$K_{\rm orb}$, 
Below, we separate $K_{\rm DOS}$ and  $K_{\rm orb}$.

  $(1/T_{1}T)_{\rm DOS}$ and $K_{\rm DOS}$ should obey the Korringa relation,
\begin{equation}
\frac{1}{(T_1 T)_{\rm DOS } }= \frac{1}{S}\frac{4\pi k_ {\rm B}  }{\hbar} (\frac{\gamma_{\rm n}}{\gamma_{\rm e}})^2 K_{\rm DOS }^2
\label{Korringa}
\end{equation}
where $\gamma_{n,e}$ is the nuclear (electron) gyromagnetic ratio, and  $S$ = 1 in the original Korringa theory \cite{KorringaR}.
From this we obtained $K_{\rm DOS}$ = 0.12\%.  
There  exist many sources that make  $S$ deviate from 1 \cite{Benett}, including an anisotropy of $g$-factor which is expected to be small for Cr-based compounds though. Therefore, we  estimate the uncertainty for  $K_{\rm DOS}$ by allowing a 20\% uncertainty for $S$.
If we adopt $S$ = 1.2 or 0.8 to estimate  the errors, the upper and lower bound for $K_{\rm DOS}^{c}$ are 0.14\% and 0.11\%, respectively.
We thus obtain 
$K_{\rm orb}^{c}$ = 0.27\%(+0.01\%/-0.02\%).
By the same manner,  $K^{\perp c}_{\rm orb} = 0.09\%$(+0.01\%/-0.02\%) is obtained. 

\begin{figure}[htbp]
\includegraphics[width=17cm]{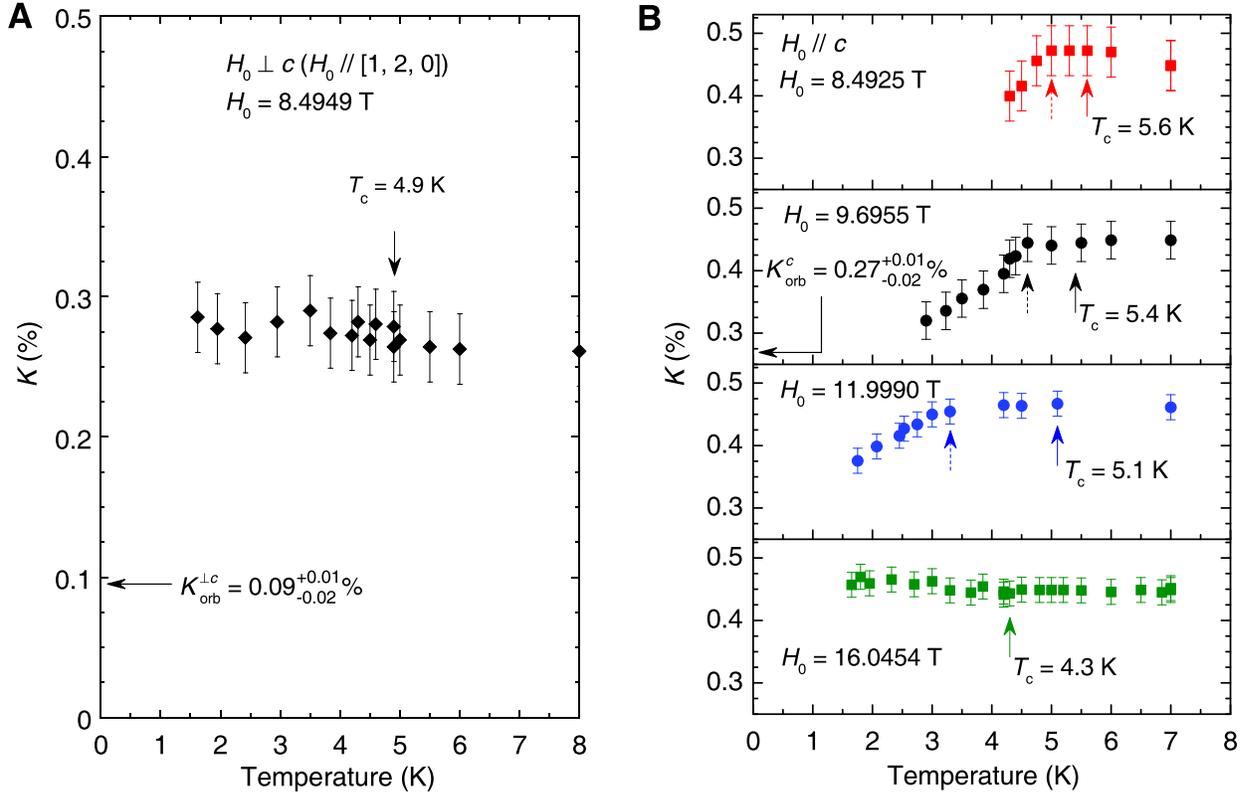}
\centering
\caption{\textbf{Temperature dependence of the Knight shift in the superconducting state.} (\textbf{A}) The temperature dependence of the Knight shift with the magnetic
field  in the $ab$ plane ($H_{0}$ $||$ [1,2,0]) 
(\textbf{B}) The Knight shift with  $H_{\rm 0} \parallel$ $c$ axis. The solid arrows indicate $T_{\rm c}$ and the dotted arrows point to the temperature $T^{*}$ below which the Knight shift starts to drop.
The error bar for $K$
was estimated by assuming that the spectrum-peak uncertainty equals the
point (frequency) interval in measuring the NMR spectra.
\label{KSC}}
\end{figure}
\begin{figure}[htbp]
	\includegraphics[width=10cm]{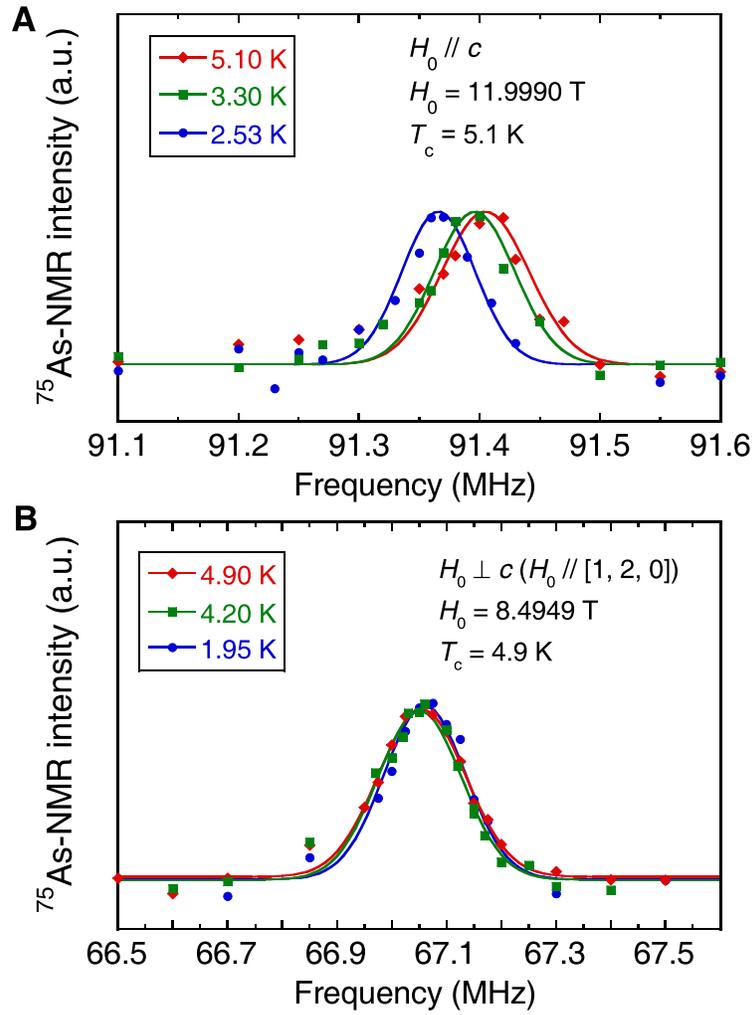}
	\centering
	\caption{\textbf{Temperature dependence of the NMR spectra for K$_{2}$Cr$_{3}$As$_{3}$ below $T_{\rm c}$.}
		(\textbf{A, B}) The $^{75}$As NMR spectra for As2 site at representative temperatures, with the magnetic field  applied parallel to the $c$ axis and in the $ab$ plane ($H_{0}$ $||$ [1,2,0]) 
		 direction, respectively. The solid curves are Gaussian function fittings to the spectra.
		\label{spectraSC}}
\end{figure}

{\bf Spin susceptibility in the superconducting state.}
Now we present the main findings of this work, namely, the spin susceptibility in the superconducting state.
%
Detailed measurements reveal that 
 $K^{\perp c}_{\rm s}$ for $H_{\rm 0}$  $||$ $ab$ and $K^{c}_{\rm s}$ for $H_{\rm 0}$ $||$  $c$ axis show very different behavior in the superconducting state, in contrast to $1/T_{\rm 1}$ which drops clearly below $T_{\rm c}$ for both field directions (Fig. S5).
 As shown in Fig.~\ref{KSC},
  $K^{\perp c}_{\rm s}$ 
 does not decrease upon cooling through the superconducting transition down to the lowest temperature measured, while $K^{c}_{\rm s}$  is  reduced significantly 
at low temperatures and vanishes toward $T$ = 0.
To appreciate more visibly the anisotropic variation of $K$, we show in
Figure \ref{spectraSC}A and B  typical spectra in the superconducting state for $H_{\rm 0}$ along the $c$ axis and in the $ab$ plane along the [1,2,0] (mirror-plane) direction, respectively.
There, it can be seen  that, the spectrum remains almost unchanged below $T_{\rm c}$ = 4.9 K for $H_{\rm 0}$ $||$ $ab$, but clearly shifts to a lower frequency below $T_{\rm c}$ = 5.1 K for $H_{\rm 0}$ $||$ $c$ axis.
Notice that, for a 
  spin-singlet superconductor, the spin
susceptibility  decreases in all directions 
and vanishes at zero temperature, and that even an inclusion of a strong spin-orbit coupling cannot account for the anisotropic reduction of the Knight shift \cite{Li2Pt3B}.  Also, the invariant Knight shift for $H_{\rm 0}$  $||$ $ab$ cannot be attributed to a pair-breaking effect due to a magnetic field as the upper critical field is even larger for this field configuration.
However,
 Cooper pairs with spin-triplet pairing have internal degrees of freedom and  the spin susceptibility below $T_{\rm c}$ can stay unchanged for some directions but is reduced along a certain direction.

\section*{Discussion}

The $\vec{d}(\vec{k})$ vector  is widely adopted to describe the order parameter 
of a spin-triplet superconducting state~\cite{SrRuO,Balian}, which  is perpendicular to the spins that comprise a   Cooper pair and  behaves like a rotation vector  in spin space.
For $H_{\rm 0}$ $||$  $\vec{d}(\vec{k})$,
$K_{\rm s}$ is reduced below $T_{\rm c}$,   while it is unchanged for $H_{\rm 0}$ $\perp$  $\vec{d}(\vec{k})$.  In superfluid $^{3}$He, there is no crystal lattice, hence  $\vec{d}(\vec{k})$ vector can rotate freely so that spin rotation symmetry is preserved~\cite{Leggett}. In solid spin-triplet superconductors,  $\vec{d}(\vec{k})$ vector is usually along a certain crystal axis so that spin rotation symmetry is spontaneously broken (spin nematicity emerges spontaneously).
In the presence of crystal disorder and spin-orbit coupling, $\vec{d}(\vec{k})$ vector
can further be pinned to a particular direction among multiple equivalent crystal axes ~\cite{Matano}. 

\begin{figure}[htbp]
\includegraphics[width=10cm]{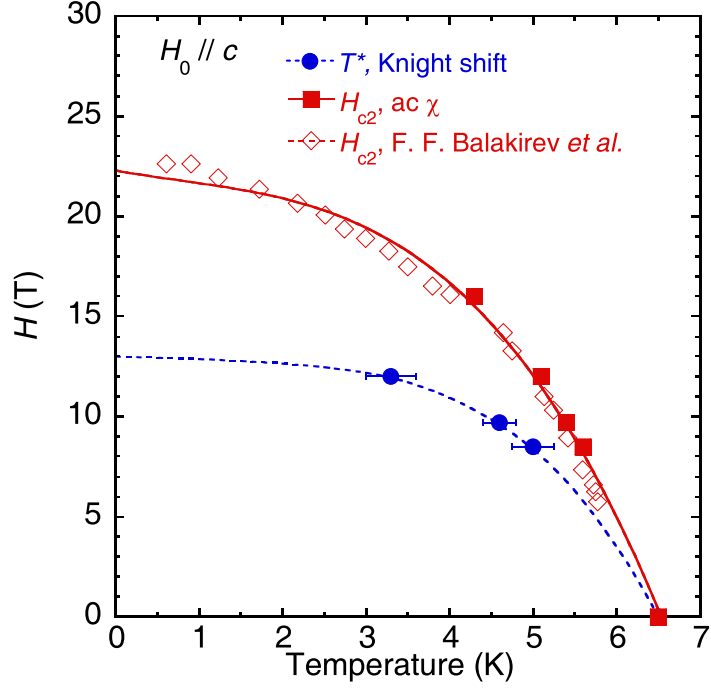}
\centering
\caption{\textbf{The $H$-$T$ phase diagram of K$_{2}$Cr$_{3}$As$_{3}$ with magnetic field along the $c$-axis}. The upper critical field data are obtained by ac susceptibility in this work (red square) and taken from ref. \cite{Canfield} (red diamond). $T^{*}$ is the temperature at which the Knight shift starts to drop. The error bar was estimated by assuming that the uncertainty equals the
	point (temperature) interval around the position indicated by the broken arrow in Fig. \ref{KSC}B. The solid and dashed curves are guides to the eyes. Below the dashed curve, the $\vec{d}(\vec{k})$ vector  is parrallel to c axis. Between the solid and dashed curves, the $\vec{d}(\vec{k})$ vector  is ascribed to  become perpendicular to c axis  (see text).
\label{Hc2}}
\end{figure}

Therefore, our results indicate that  Cooper pairs in K$_{2}$Cr$_{3}$As$_{3}$ are in a  spin-triplet state, with
the $\vec{d}(\vec{k})$ vector  along the  $c$ axis.
Such spin-triplet state possesses internal degrees of freedom and will  provide a good opportunity to explore novel phenomena such as  collective modes of the order parameter, half-quantum vortices, etc.
An exotic feature seen from Fig.~\ref{KSC}B is that $K^{c}_{\rm s}$ starts to drop at a temperature $T^{*}$ that is lower than $T_{\rm c}$.
It is emphasized that both the NMR intensity and $1/T_{\rm 1}$ drop sharply at $T_{\rm c}$ (Fig. S4, S5). In particular, the former quantity is measured under exactly the same condition as $K$, which assure that the measured  $T_{\rm c}$ represents the intrinsic superconducting transition temperature.
 The temperature difference between $T^{*}$ and $T_{\rm c}$ increases with increasing magnetic field, and $K^{c}_{\rm s}$ even shows no  reduction for $H_{\rm 0}$ = 16 T although this field is smaller than $H_{\rm c2}$. Figure~\ref{Hc2} shows  $T^{*}$ and $T_{\rm c}$($H$) obtained under different fields.
  There is no evidence showing another phase transition in this temperature range 
  from the electromagnetic, heat transport measurements or our NMR spectra. 
  Therefore, the $H-T$ phase diagram of Fig. ~\ref{Hc2} is ascribed  to a unlocking of the $\vec{d}(\vec{k})$ vector by the magnetic field. The curve shown by the broken lines represents the pinning force in terms of field (pinning field ) $H^*$  above which Zeeman energy wins so that the $\vec{d}(\vec{k})$ vector originally pinned to the $c$-axis direction is unlocked  and rotates 90 degrees. 
  Recall that the $\vec{d}(\vec{k})$ vector is perpendicular to the Cooper-pair spins.
   $H^*$ is no larger than 13 T.
  Also note that such $\vec{d}(\vec{k})$ vector depinning to gain Zeeman energy is different from the $\vec{d}(\vec{k})$-vector rotation between two nearly-degenate states \cite{Kim}. 
  
  In passing, we make two comments. 
 First, a tiny change of the Knight shift was found below $T_{\rm c}$ along specific crystal directions of a strongly-correlated material UPt$_3$ \cite{UPt3NMR}, but the interpretation of the result is controvertial \cite{UPt3Review}, as the change is less than 1\% of the total Knight shift. 
 Second,  inversion symmetry is broken in  K$_2$Cr$_3$As$_3$ so that   parity mixing can occur. However, the band splitting due to  inversion-symmetry breaking  is about 60 meV \cite{Jiang}, which is comparable to all spin-singlet noncentrosymmetric superconductors including Li$_2$Pd$_3$B \cite{Li2Pd3B}. Therefore,  parity mixing should be small in the present case. In fact, if a parity mixing takes place, it is the singlet component that is mixed. Then, one should see a certain decrease in the Knight shift even for $H\parallel ab$. However, we do not observe such behavior.
 

	Finally, we discuss the orbital wave function of the Cooper pairs in K$_{2}$Cr$_{3}$As$_{3}$.
	Density function theory  calculations show that there
	are three bands across the Fermi level, namely, two quasi
	one-dimensional (1D) bands $\alpha$ and $\beta$, and one three-dimensional
	(3D) band $\gamma$ \cite{Jiang}.  
	The  $\gamma$ band makes the
	dominant contribution (75\%) to the density of states.
	 Previous spin-relaxation rate study has revealed point nodes in the gap function ~\cite{Yang,LuoPRL}.
For a 3D Fermi surface, 
group theory analysis shows that in the spin-triplet pairing channel, gap functions with both 
point nodes and line nodes are allowed (see SM). In the case of point nodal gap, 
 all the point nodes are located at the two poles on
the Fermi surface with $k_x$ = $k_y$ = 0, as  
listed in Table I.
Among them, only  $E'$ states ($p_x$+$ip_y$ and $p_x$-$ip_y$) are
  consistent with our Knight shift result with the quantum axis along the $c$ axis direction.
 Notice that, for all  possible $E'$ states which are linear combinations of two basis functions, the two states listed in  Table I are energetically favored, since the $\vec{d}(\vec{k})$-vector is along the $\vec{z}$ direction.  \cite{Venderbos,Yao}.
Such a state breaks time reversal symmetry and is consistent with zero-field $\mu$SR measurement that revealed evidence for a spontaneous appearance of a weak internal magnetic field below $T_{\rm c}$ \cite{usr}.

An  $E'$ state is analogous to the A phase (or Anderson-Brinkman-Morel state) 
in superfluid $^{3}$He \cite{Leggett}, and was initially proposed 
as a superconducting state for Sr$_2$RuO$_4$ \cite{Rice} but not supported by the recent experiment \cite{Brown}.
Such state is topological, therefore 
Majorana zero modes can be expected in vortex cores \cite{Ivanov,Volovik}.
In particular, if a superconducting thin film of K$_2$Cr$_3$As$_3$, with its thickness  smaller than the superconducting coherence length, is available,  a single Majorana zero mode will be expected in the core of a half-quantum vortex.
Thus, our results demonstrate that K$_{2}$Cr$_{3}$As$_{3}$  is a new platform for basic research of topological materials  and possible technical applications of topological superconductivity. We also hope that our work will stimulate more precise measurements using single crystals to look for novel phenomena arising from the internal degrees of freedom of spin triplet pairing, including multiple phases and those aforementioned.


\section*{Materials and methods}

 \textbf{Sample preparation}

High quality single crystal  K$_{2}$Cr$_{3}$As$_{3}$ samples used in this work were grown by self-flux method as described in Ref.~\cite{K}. First, the starting materials KAs and CrAs were prepared by reacting K pieces, Cr powder and As powder. The mixture of KAs and CrAs with a  molar ratio of 6:1  was placed in an alumina crucible and sealed in evacuated Ta crucible and  quartz tube. They were then sintered at 1273 K for 24 h, followed by cooling down at 1 K/h. Extra flux was removed to obtain single crystals by centrifugation at 923 K. The single crystals are straight, thin and needle-like, with typical length of 5 mm and diameter of tens of microns. The $c$ axis of the crystal is easy to recognize, which is along the direction of the  needle.
The sample quality was checked  by dc  susceptibility, which shows $T_{\rm c}$ $\approx$ 6.5 K at zero field. During the NMR experiments, $T_{\rm c}$  was  confirmed  by measuring the inductance of the NMR coil. The sharp decrease of $1/T_{1}$  below $T_{\rm c}$ further ensures the high sample quality.


\textbf{NMR measurements}

For NMR measurements with the magnetic field parallel to the $ab$ plane, only one needle was used.
For NMR measurements along the $c$ axis, several needles were selected  and aligned together.
Because the K$_{2}$Cr$_{3}$As$_{3}$ sample is fragile and air-sensitive, the   sample handling was done in an Ar-protected glove box.
The $^{75}$As (nuclear spin \textit{I} = 3/2 with  nuclear gyromagnetic ratio $\gamma$ = 7.2919 MHz/T) NMR measurements were carried out by using a phase-coherent spectrometer. The NMR spectra were obtained by scanning the frequency point by point and integrating the spin echo at a fixed magnetic field $H_{0}$. The spin echo was observed by using a standard $\pi/2-\tau-\pi$ pulse sequence with $\pi/2$ pulse length of 7 $\mu$s and $\tau$ = 40 $\mu$s. 
An Attocube piezo horizontal rotator was used for angle-variated NMR measurements of $H_{\rm 0}$  $\parallel$  $ab$ plane.  The angle repeatability is 50 m$^{\circ}$ and the resolution is 6 m$^{\circ}$ for the rotator. Two orthogonal Hall bars were placed on the sample holder to check the field orientation and to ensure  the rotation axis being perpendicular to the applied magnetic field.
The  spin-lattice relaxation rate $1/T_{1}$ was measured by the saturation-recovery method,
and  determined by a good fitting  of the nuclear magnetization to $1-M(t)/M(\infty) = 0.1exp(-t/T_{1}) + 0.9exp(-6t/T_{1})$, where $M(\infty)$ and $M(t)$ are the nuclear magnetization in the thermal equilibrium and at a time $t$ after the saturating pulse, respectively.

\bibliography{}

\begin{thebibliography}{99}

\bibitem{BCS}
	J.  L. Bardeen, N. Cooper, J. R. Schrieffer,
	Theory of Superconductivity. \emph{Phys. Rev.} {\bf 108},
	1175 (1957).
	
		
\bibitem{cuprates}
	C. C. Tsuei, J. R. Kirtley, Pairing symmetry in cuprate superconductors.
	\emph{Rev. Mod. Phys.} {\bf 72}, 969 (2000).
	
\bibitem{Leggett}
	A. J. Leggett,
	A theoretical description of the new phases of liquid $^{3}$He.
	\emph{Rev. Mod. Phys.} \textbf{47}, 331 (1975).

	
\bibitem{neutronstars}
	D. Page, M. Prakash, J. M. Lattimer,  A. W. Steiner, Rapid Cooling of the Neutron Star in Cassiopeia A Triggered	by Neutron Superfluidity in Dense Matter. \emph{Phys. Rev. Lett.} \textbf{106}, 081101 (2011).
	

\bibitem{SrRuO}
	Y. Maeno, H. Hashimoto, K. Yoshida, S. Nishizaki, T. Fujita, J. G. Bedonorz, F. Lichtenberg, Superconductivity in a layered perovskite without copper.
	\emph{Nature} \textbf{372}, 532 (1994).
	

	
	
	
	
	

\bibitem{UPt3Review}
	R. Joynt, L. Taillefer,
	The superconducting phases of UPt$_3$.
	\emph{Rev. Mod. Phys}. \textbf{74}, 235 (2002).
	
	
	
	
	
	\bibitem{UGe2}	
	S. S. Saxena, P. Agarwal, K. Ahilan, F. M. Grosche, R. K. W. Haselwimmer, M. J. Steiner, E. Pugh, I. R. Walker, S. R. Julian, P. Monthoux, G. G. Lonzarich, A. Huxley, I. Sheikin, D. Braithwaite, J. Flouquet,
Superconductivity on the border of itinerant-electron ferromagnetism in UGe$_2$.	
	\emph{Nature} {\bf 406}, 587 (2000).
	
	\bibitem{URhGe}
	D. Aoki, A. Huxley, E. Ressouche, D. Braithwaite,J. Flouquet, J.-P. Brison, E. Lhotel, C. Paulsen,
	Coexistence of superconductivity and ferromagnetism in URhGe.
	\emph{Nature} {\bf 416}, 613 (2001).
	
	\bibitem{UCoGe}
	N. T. Huy, A. Gasparini, D. E. de Nijs, Y. Huang, J. C.	P. Klaasse, T. Gortenmulder, A. de Visser, A. Hamann, T. Gorlach, H. von Lohneysen,
	Superconductivity on the Border of Weak Itinerant Ferromagnetism in UCoGe.
	\emph{Phys. Rev. Lett.} {\bf 99}, 067006 (2007).
	
	\bibitem{UTe2}
	S. Ran, C. Eckberg, Q-P. Ding, Y. Furukawa, T. Metz, S. R. Saha, I-L. Liu, M. Zic, H. Kim, J. Paglione, N. P.	Butch,
	Nearly ferromagnetic spin-triplet superconductivity.
	\emph{Science} {\bf 365}, 684 (2019).
	
	
	


	
\bibitem{Rice}
	T. M. Rice, M. Sigrist,
	Sr$_{2}$RuO$_{4}$: an electronic analogue of $^{3}$He?
	\emph{J. Phys. Condens. Matter} \textbf{7}, L643 (1995).
	
	
\bibitem{Brown}
	A. Pustogow,
	Yongkang Luo,
	A. Chronister,
	Y.-S. Su,
	D. A. Sokolov,
	 F. Jerzembeck,
	A. P. Mackenzie,
	C. W. Hicks,
	N. Kikugawa,
	S. Raghu,
	E. D. Bauer,
	S. E. Brown,
	Constraints on the superconducting order parameter in Sr$_{2}$RuO$_{4}$ from oxygen-17 nuclear magnetic resonance.
	\emph{Nature} \textbf{574}, 72 (2019).
	
	
	\bibitem{Fu_Berg_PhysRevLett.105.097001}
	L. Fu, E. Berg,
	Odd-Parity Topological Superconductors: Theory and Application to Cu$_x$Bi$_2$Se$_3$. \textit{Phys. Rev. Lett.} \textbf{105}, 097001 (2010).
	
	\bibitem{Matano}
	K. Matano,
	M. Kriener,
	K. Segawa,
	Y. Ando,
	Guo-qing Zheng,
	Spin-rotation symmetry breaking in the superconducting state of Cu$_{x}$Bi$_{2}$Se$_{3}$.
	\emph{Nat. Phys.} \textbf{12}, 852 (2016).
	


\bibitem{Wilczek}
	F. Wilczek, Majorana returns. \emph{Nat. Phys.} \textbf{5}, 614 (2009).


	\bibitem{Qi}
	X. L. Qi,
	S. C. Zhang,
	Topological insulators and superconductors.
	\emph{Rev. Mod. Phys.} \textbf{83}, 1057 (2011).

	
	
\bibitem{Kitaev}
	A. Y. Kitaev,
	Unpaired Majorana fermions in quantum wires.
	\emph{Physics-Uspekhi} \textbf{44}, 131 (2001).
	
	\bibitem{Fu}
	L. Fu,
	C. L. Kane,
	Superconducting proximity effect and Majorana fermions at the surface of a topological insulator.
	\emph{Phys. Rev. Lett.} \textbf{100}, 096407 (2008).
	
\bibitem{Mourik}
	V.  Mourik,
	K. Zuo,
	S. M. Frolov,
	S. R. Plissard,
	E. P. A. M. Bakkers,
	 L. P. Kouwenhoven,
	Signatures of Majorana fermions in hybrid superconductor-semiconductor nanowire devices.
	\emph{Science} \textbf{336}, 1003 (2012).

	
\bibitem{Jiajinfeng}
	H. H. Sun,
	K. W. Zhang,
	L. H. Hu,
	C. Li,
	G. Y. Wang,
	H. Y. Ma,
	Z. A. Xu,
	C. L. Gao,
	 D. D. Guan,
	 Y. Y. Li,
	C. H. Liu,
	D.  Qian,
	Y. Zhou,
	L. Fu,
	S. C. Li,
	F. C. Zhang,
	J. F. Jia,
	Majorana Zero Mode Detected with Spin Selective Andreev Reflection in the Vortex of a Topological Superconductor.
	\emph{Phys. Rev. Lett.} \textbf{116}, 257003 (2016).
	
	\bibitem{Ding}
	D. Wang, L. Kong, P. Fan, H. Chen, S. Zhu, W. Liu, L. Cao, Y. Sun, S. Du, J. Schneeloch, R. Zhong, G. Gu, L. Fu, H. Ding, H.-J. Gao,
	\emph{Science} {\bf 362}, 333 (2018).
	
	
	
	
	\bibitem{Senthil}
	T. Senthil, J. B. Marston, M. P. A. Fisher, Spin quantum Hall effect in unconventional superconductors. \textit{Phys. Rev. B} \textbf{60}, 4245-4254 (1999).
	
	
	
	
	\bibitem{sato_PhysRevB.79.094504}
	M. Sato, S. Fujimoto, Topological phases of noncentrosymmetric superconductors: Edge states, Majorana fermions, and non-Abelian statistics. \textit{Phys. Rev. B} \textbf{79}, 094504 (2009).
	
	\bibitem{Li2Pt3B}
	M. Nishiyama, Y. Inada,  G.-q. Zheng,
	Spin Triplet Superconducting State due to Broken Inversion Symmetry in Li$_2$Pt$_3$B.
	\textit{Phys. Rev. Lett.} \textbf{98}, 047002 (2007).
	
\bibitem{K}
	J. K. Bao,
	J. Y. Liu,
	C. W. Ma,
	Z. H. Meng,
	Z. T. Tang,
	Y. L. Sun,
	H. F. Zhai,
	H. Jiang,
	H. Bai,
	C. M. Feng,
	Z. A. Xu,
	G. H. Cao,
	Superconductivity in Quasi-One-Dimensional K$_{2}$Cr$_{3}$As$_{3}$ with Significant Electron Correlations.
	\emph{Phys. Rev. X} \textbf{5}, 011013 (2015).	

	
	\bibitem{Rb}
	Z. T. Tang,
	J. K. Bao,
	Y. Liu,
	Y. L. Sun,
	A. Ablimit,
	H. F. Zhai,
	H. Jiang,
	C. M. Feng,
	Z. A. Xu,
	G. H. Cao,
	Unconventional superconductivity in quasi-one-dimensional Rb$_{2}$Cr$_{3}$As$_{3}$.
	\emph{Phys. Rev. B} \textbf{91}, 020506 (2015).
	
	\bibitem{Cs}
	Z. T. Tang,
	J. K. Bao,
	Z. Wang,
	H. Bai,
	H. Jiang,
	Y. Liu,
	H. F. Zhai,
	C. M. Feng,
	Z. A. Xu,
	G. H. Cao,
	Superconductivity in quasi-one-dimensional Cs$_{2}$Cr$_{3}$As$_{3}$ with large interchain distance.
	\emph{Sci. China Mater.} \textbf{58}, 16 (2015).
	
	\bibitem{Na}
	Q. G. Mu,
	B. B. Ruan,
	B. J. Pan,
	T. Liu,
	J. Yu,
	K. Zhao,
	G. F. Chen,
	Z. A. Ren,
	Ion-exchange synthesis and superconductivity at 8.6 K of $\mathrm{N}{\mathrm{a}}_{2}\mathrm{C}{\mathrm{r}}_{3}\mathrm{A}{\mathrm{s}}_{3}$ with quasi-one-dimensional crystal structure.
	\emph{Phys. Rev. Mater.} \textbf{2}, 034803 (2018).
	
	\bibitem{Yang}
	J. Yang,
	Z. T. Tang,
	 G. H. Cao,
	Guo-qing Zheng,
	Ferromagnetic Spin Fluctuation and Unconventional Superconductivity in Rb$_{2}$Cr$_{3}$As$_{3}$ Revealed by $^{75}$As NMR and NQR.
	\emph{Phys. Rev. Lett.} \textbf{115}, 147002 (2015).
	
	\bibitem{LuoPRL}
	J. Luo,
	J. Yang,
	R. Zhou,
	Q. G. Mu,
	T. Liu,
	Z. A. Ren,
	C. J. Yi,
	Y. G. Shi,
	Guo-qing Zheng,
	Tuning the Distance to a Possible Ferromagnetic Quantum Critical Point in A$_{2}$Cr$_{3}$As$_{3}$.
	\emph{Phys. Rev. Lett.} \textbf{123}, 047001 (2019).
	
	
	\bibitem{Imai}
	H. Z. Zhi,
	T. Imai,
	F. L. Ning,
	J. K. Bao,
	G. H. Cao,
	NMR Investigation of the Quasi-One-Dimensional Superconductor K$_{2}$Cr$_{3}$As$_{3}$.
	\emph{Phys. Rev. Lett.} \textbf{114}, 147004 (2015).
	
	
	
	
	\bibitem{Canfield}
	F. F. Balakirev,
	T. Kong,
M. Jaime,
		R. D. McDonald,
	 C. H. Mielke,
	A. Gurevich,
	P. C. Canfield,
	S. L. Bud'ko,
	Anisotropy reversal of the upper critical field at low temperatures and spin-locked superconductivity in K$_{2}$Cr$_{3}$As$_{3}$.
	\emph{Phys. Rev. B} \textbf{91}, 220505(R) (2015).
	
	
	\bibitem{Yuan_HQ}
	G. M. Pang,
	M. Smidman,
	W. B. Jiang,
	J. K. Bao,
	Z. F. Weng,
	Y. F. Wang,
	L. Jiao,
	J. L. Zhang,
	G. H. Cao,
	H. Q. Yuan,
	Evidence for nodal superconductivity in quasi-one-dimensional K$_{2}$Cr$_{3}$As$_{3}$.
	\emph{Phys. Rev. B} \textbf{91}, 220502(R) (2015).
	
	
	
	
	\bibitem{usr}
D. T. Adroja,
	A. Bhattacharyya,
	M. Telling,
	Y.  Feng,
	M. Smidman,
	B. Pan,
	J. Zhao,
	A. D. Hillier,
	F. L. Pratt,
	A. M. Strydom,
	Superconducting ground state of quasi-one-dimensional K$_{2}$Cr$_{3}$As$_{3}$ investigated using $\mu$SR measurements.
	\emph{Phys. Rev. B} \textbf{92}, 134505 (2015).
	
	
	
	
	
	
	
	
	
	
	
	
	
	
	
\bibitem{Abragam}
A. Abragam, \emph{The Principles of Nuclear Magnetism}, Oxford University Press, London, (1961).
	
\bibitem{LuoCPB}
	J. Luo, C. G. Wang, Z. C. Wang, Q. Guo, J. Yang, R. Zhou, K. Matano, T. Oguchi, Z. A. Ren, G. H. Cao, Guo-qing Zheng,
	NMR and NQR studies on transition-metal arsenide superconductors LaRu$_{2}$As$_{2}$, KCa$_{2}$Fe$_{4}$As$_{4}$F$_{2}$ and A$_{2}$Cr$_{3}$As$_{3}$.
	\emph{Chinese Physics B}, \textbf{29}, 067402 (2020).


\bibitem{Moriya}
T. Moriya,
\emph{Spin Fluctuations in Itinerant Electron Magnetism}. Springer-Verlag, Berlin, 1985.

\bibitem{KorringaR}
J. Korringa,
Nuclear magnetic relaxation and resonance line shift in metals.
\emph{Physica} \textbf{16}, 601 (1950).



\bibitem{Benett}
G. C. Carter, L. H. Bennet, D. J. Kahn,
\emph{Mettalic shifts in NMR}. Pergamon, 1977.



	\bibitem{Balian}
R. Balian, N. R. Werthamer,
Superconductivity with Pairs in a relative p wave.
\emph{Phys. Rev.} \textbf{131}, 1553 (1963).






\bibitem{Kim}
B. Kim, S. Khmelevskyi, I. I. Mazin, D. F. Agterberg, C. Franchini,
Anisotropy of magnetic interactions and symmetry of the order
parameter in unconventional superconductor Sr$_2$RuO$_4$.
\textit{npj Quantum Materials}  \textbf{2},37 (2017).
	
\bibitem{UPt3NMR}
H. Tou, Y. Kitaoka, K. Ishida, K. Asayama, N. Kimura, Y. Onuki, E. Yamamoto, Y. Haga, K. Maezawa,
Nonunitary spin-triplet superconductivity in UPt$_{3}$: evidence from $^{195}$Pt knight shift study.
\emph{Phy. Rev. Lett.} \textbf{80}, 3129 (1998).
	
	
	\bibitem{Jiang}
	H. Jiang, G. H. Cao, C. Cao,
	Electronic structure of
	quasione-dimensional superconductor K$_{2}$Cr$_{3}$As$_{3}$ from first principles
	calculations. \emph{Sci. Rep}. {\bf 5}, 16054 (2015).
	
	\bibitem{Li2Pd3B}	
	M. Nishiyama, Y. Inada, and G.-q. Zheng, Superconductivity of the ternary boride Li2Pd3B
	probed by $^{11}$B NMR, 
	\emph{Phys. Rev. B} \textbf{71}, 220505(R) (2005).
	
		\bibitem{Venderbos}
	J. W. F. Venderbos, V. Kozii, L. Fu,
	Odd-parity superconductors with two-component order parameters: Nematic and chiral, full gap, and Majorana node.
	\emph{Phys. Rev. B}  {\bf 94}, 180504(R) (2016).
	
	\bibitem{Yao}
	W. Huang,  H. Yao,
	Possible Three-Dimensional Nematic Odd-Parity Superconductivity in Sr$_{2}$RuO$_{4}$.
	\emph{Phys. Rev. Lett.} {\bf 121}, 157002 (2018).
	
	
	\bibitem{Ivanov}
	 D. A. Ivanov,
	Non-Abelian Statistics of Half-Quantum Vortices in $p$-Wave Superconductors.
	\emph{Phys. Rev. Lett.}  {\bf 86}, 268 (2001).
	
	\bibitem{Volovik}	
	G. E. Volovik, {\it The Universe in a Helium Droplet}, Oxford University Press, Oxford, 2002.
	

	
\bibitem{deGennes}
P. G. de Gennes,
\emph{Superconductivity of Metals and Alloys}. Westview Press, Oxford, 1999.




\bibitem{RRR}
G. H. Cao, J. K. Bao,  Z. T. Tang, Y. Liu, H. Jiang,
Peculiar properties of Cr$_{3}$As$_{3}$-chain-based superconductors.
\emph{Philosophical Magazine} {\bf 97}, 591 (2017).



\bibitem{Cu0.5Bi2Se3}
T. Kawai, C. G. Wang, Y. Kandori, Y. Honoki, K. Matano, T. Kambe, Guo-qing Zheng,
Direction and symmetry transition of the vector order parameter in topological superconductors Cu$_x$Bi$_2$Se$_3$.
\emph{Nat. Commun.} \textbf{11}, 235 (2020)

\bibitem{YZ17}
Y. Zhou, C. Cao,  F. C. Zhang,
Theory for superconductivity in alkali chromium arsenides  A$_{2}$Cr$_{3}$As$_{3}$ (A = K, Rb, Cs).
\emph{Science Bulletin} \textbf{62}, 208 (2017).




\end{thebibliography}
\bibliographystyle{Science}

\section*{Acknowledgments}
We thank G.H. Cao, J.P. Hu, K. Ishida, Y. Tanaka, T. Xiang and R. Zhou for interests and useful discussion.

\textbf{Funding:} This work was supported by the National Key Research and Development Program of China (Nos. 2017YFA0302904, 2017YFA0302901 and 2016YFA0300502), the National Natural Science Foundation of China (Nos. 11634015, 11674377, 11774306, and 12034004), 
 K. C. Wong Education Foundation (No. GJTD-2018-01), the Youth Innovation Promotion Association of CAS (No. 2018012), as well as JSPS (No. JP19H00657). 
 
 \textbf{Author contributions:} G.-q.Z designed and coordinated the project. C.J.Y and Y.G.S synthesized  the  single crystals. 
J.Y and J.L  performed  NMR and other measurements. Y.Z conducted group theory analysis. G.-q.Z wrote the manuscript with inputs from  J.Y and Y.Z.  All authors discussed the results and interpretation.  

\textbf{Competing interests:} The authors declare no competing interests. 

\textbf{Data and materials availability:} All data needed to evaluate the conclusions in the paper are
present in the paper and/or the Supplementary Materials. 



\clearpage

\clearpage

\clearpage

\clearpage

\clearpage

\clearpage

\clearpage

\clearpage
\begin{table}[h]
	\centering
	\caption{All the possible superconducting gap functions
		that give rise to spin triplet and point nodes on a $D_{3h}$ lattice.}
	
\begin{tabular}{c|c}
	\hline\hline
	$\Gamma$ &spin-triplet $\vec{d}(k)$\\ 
	\hline
	$E'$ & $(p_x \pm i p_y) \vec{z}$  \\
	\hline
	$A_1'$ & $p_x\vec{x} + p_y \vec{y}$  \\
	\hline
	$A_2''$ & $p_y\vec{x} - p_x \vec{y}$ \\
	\hline
	 $E''$ & ($p_x\vec{x} - p_y \vec{y}$, $p_y\vec{x} + p_x \vec{y}$) \\
	\hline\hline
\end{tabular}

\end{table}

\end{document}